\documentclass[twocolumn,prl,amssymb]{revtex4}
\usepackage{graphicx}% Include figure files
\usepackage{dcolumn}% Align table columns on decimal point
\usepackage{bm}% bold math
\usepackage{psfrag,epsfig}

\newcommand{\Opd}{{\mathcal{O}}(p^2)}

\newcommand{\be}{\begin{equation}}
\newcommand{\ee}{\end{equation}}
\newcommand{\ba}{\begin{eqnarray}}
\newcommand{\ea}{\end{eqnarray}}

\begin{document}

\title{
Reply to Comment on ``Surprises in threshold antikaon-nucleon physics"}

\author{Jos\'e A. Oller$^1$, Joaquim Prades$^2$ and Michela Verbeni$^{1}$}
\affiliation{$^1$ Departamento de F\'{\i}sica, Universidad de Murcia, 
E-30071 Murcia, Spain\\$^2$ Centro Andaluz de F\'{\i}sica
de las Part\'{\i}culas Elementales (CAFPE) and 
Departamento de F\'{\i}sica Te\'orica y del Cosmos,\\ 
Universidad de Granada, 
Campus de Fuenta Nueva, E-18002 Granada, Spain}
%\date{\today}

\begin{abstract}
\noindent
In their Comment, Borasoy {\it et al.} \cite{comment}, 
criticize our results \cite{pom}  that accommodate
both scattering data 
and the new accurate measurement by DEAR
 of the shift and width of kaonic hydrogen.
In our calculations we have employed unitary chiral perturbation theory 
(UCHPT).  We discuss why their arguments are irrelevant or do not hold.
\end{abstract}

%\pacs{36.10.Gv, 11.80.-m, 12.39.Fe, 13.75.Jz}

\maketitle

\noindent

{\bf 1.} Borasoy et al. correctly state in their Comment that
 the interacting kernel
employed in the Letter \cite{pom} can induce residual cuts 
 in some energy regions above the threshold of the lighter channels 
($\pi \Lambda$ and   $\pi \Sigma$) because of $u$-crossed 
exchange of baryons 
in amplitudes with heavier states. 
  However, these cuts only induce appreciable 
unitarity 
deviations  in $\pi^0 \Lambda\rightarrow \pi^0 \Lambda$
at energies lower than 1.4 GeV, where no data exists to compare with. 
These deviations  are negligible for the $\pi\Sigma$ and 
do not appear at all in the rest of channels. 
  UCHPT resums the right-hand cut (or unitarity cut) while the 
 contributions from the other cuts in the interacting kernel  
are calculated only at a given order in  CHPT, 
by matching with the chiral series  order by order. 
The right hand cut  resummation is justified 
since  the chain of unitarity bubbles 
 is enhanced by the large masses of kaons and baryons \cite{wein}. 
In this way, one reproduces the CHPT results
up to a given order, e.g. to ${\Opd}$ in  \cite{pom}.
In this scheme, calculating the interacting kernel
at higher orders in CHPT 
will eventually soften the effect of these spurious cuts.
If one eliminates the $u$-channel cuts
 in  the S-wave projections by ad hoc taking  them  
to be constants above some energy value,
  as in \cite{prl}, then {\it chiral symmetry as well as crossing are 
violated} even at leading order in CHPT.

{\bf 2.} Borasoy et al. \cite{comment} consider necessarily unphysical 
the narrow pole just below the $K^- p$ threshold at (1431.21-$i\,$1.28)
 MeV. First of all, one must point
 out that this pole does not occur in the $physical$ Riemann sheet 
but in the unphysical ones as required by hermiticity.
 Furthermore, the fact that this resonance turns out to be so narrow has 
a clear physical origin and is not an artifact at all. This resonance has couplings (with
respect to $I=1$ states) of (1.8,4.5,1.0,0.7,0.5) GeV to the 
($K\Xi$,$\eta\Sigma$,$\bar{K}N$,$\pi \Sigma$,$\pi \Lambda$), respectively. 
One 
could be surprised that the couplings to heavier channels 
(the $\eta \Sigma$ threshold is around 
200 MeV above the resonance mass) are larger than those to the channels closer in energy.
But this is also the case for the $\Lambda(1670)$, well known 
experimentally, that couples much more
strongly to the $K\Xi$ channel \cite{team}
than to the rest of channels. The pole at 1431 MeV is so narrow  because of
the smallness of its couplings to the open channels.
 They also argue that such narrow pole should lead to violations of
fundamental principles like the Wigner condition. 
 These authors apply such condition to the phase shifts.
However in the inelastic case (as the one here), 
 one should consider the phase of the scattering amplitude.
 This increases the derivative from -300  \cite{comment} to  
-150 fm and makes 
Figure  1 in  \cite{comment} irrelevant to the present discussion 
since inelasticity varies sharply.
 Furthermore, the Wigner condition is
not violated in the presence of a narrow pole, 
as the derivative of the phase is
positive (take a Breit-Wigner) and the Wigner condition is
a lower negative bound. 
The phase of the scattering amplitude
starts to decrease with energy between the pole and the 
opening of the $K^-p$ channel at 1431.95 MeV due to  the
presence of the branch singularity  associated
 with the opening of 
 this channel.
The Wigner condition 
cannot be applied  on top of the thresholds of new opening channels, 
simply because the phase  is not differentiable in an inelastic
 branch point and  the derivative  near it tends to  infinity.
 If the situation is softened, by varying the masses 
a few MeV so that the $K^- p$ threshold  moves away,  then
 the Wigner limit is restored.

{\bf 3.} The authors of \cite{comment} claim
 that this pole should be seen in the $\pi\Sigma$ event 
 distribution. However, this is not so because the $\pi \Sigma$ distribution 
 is dominated by $I=0$ and this narrow resonance is $I=1$.
  But 
even if we use the $K^-p \rightarrow \pi^\pm\Sigma^\mp$  amplitudes
one does not observe any narrow peak because of the  interference with $I=0$,
which is larger than $I=1$, as we have checked.
One would need  much better precision than in present data to observe the 
modifications in the event distribution due to such pole. 
Finally, the authors  of \cite{comment} 
also state that our fit is too sensitive to the 
values of the  fitted parameters. One must
 remark that the set of parameters is highly correlated so it does not make any sense to
 arbitrarily vary their values.
What is sensible is to change the set of fitted data, e.g. by modifying
 the values of 
$\sigma_{\pi N}$, $m_0$  or by considering 
new data points, and check that similar fits arise.

\end{document}